# How Semantic Information G Measure Relates to Distortion, Freshness, Purposiveness, and Efficiency

Chenguang Lu

*Abstract*—To improve communication efficiency and provide more useful information, we need to measure semantic information by combining inaccuracy or distortion, freshness, purposiveness, and efficiency. The author proposed the semantic information G measure before. This measure is more compatible with Shannon information theory than other semantic or generalized information measures and has been applied to machine learning. This paper focuses on semantic predictive information (including observation information) and purposive (or goal-related) information (involving semantic communication and constraint control). The GPS pointer is used as an example to discuss the change of semantic predictive information with inaccuracy and time (age of the information). An example of constraint control (controlling probability distributions) is provided for measuring purposive information and maximizing this information and the information efficiency. The information rate fidelity function (a generalization of the information rate distortion function) is introduced for the optimization. Two computing examples demonstrate how to measure predictive and goal-related information and optimizing information efficiency. The results accord with theoretical conclusions well. The G measure is related to deep learning; its application to machine learning is worth exploring. Communication efficiency also involves utilities or information values; semantic communication optimization combining utilities needs further research.

**Keywords: Semantic information measure, Distortion, information freshness, goal-related information, communication efficiency.**

## I. INTRODUCTION

The internet provides a large amount of information, but for specific recipients, most information conveyed by bits is useless or inefficient because 1) people receive information according to semantic meaning instead of probability's change; 2) some information is inaccurate or wrong, some information is outdated, and some information seems new but without new meaning; 3) only information that is related to people's needs (including the needs of work, life, knowledge, aesthetic feeling, and curiosity) is useful. Therefore, it is important to improve communication efficiency, using less Shannon information [1] to convey more semantic or useful information.

To improve communication efficiencies, researchers have made significant progress related to distortion [2], [3], [4], information freshness [5], [6], [7], [8], information values [9], [10], [11], purposiveness [12], [13], and so on. In addition, more and more researchers have realized the importance of semantic communication [9], [14], especially for machine learning and the future internet (including the internet of things) [12], [13].

To this end, we need a proper semantic information measure [2] (involving logical probability [15], truth and falsehood [16], and distortion or inaccuracy [17]) with the considerations of information's freshness (or timeliness) and purposiveness (about whether actual results, i.e., probability distributions produced by constraint control, accord with the corresponding purposes or goals) [13], [18], [19]. We also need ways to measure and maximize communication efficiencies, which come from comparing Shannon information or bits with semantic information, fidelity or distortion [1], [20], cost [21], energy [22], and utility or value [9], [10]. This paper only considers information efficiency, for which we use less Shannon information to convey more semantic information with less distortion (note "information efficiency" in economics or the efficient market theory has a different meaning). Communication efficiencies include information efficiency. For information efficiency, Shannon proposed the information rate distortion function $R(D)$ [2] to tell the Minimum Mutual Information (MMI) $R$ for the given upper limit $D$ of distortion. We need a similar function to tell the MMI $R$ for the given lower limit of average semantic information.

Researchers have proposed many information measures for semantic information [15], [23], [24], [25], [26], [27], generalized information [28], [29], and fuzzy information [30], [31], [32]. I proposed the semantic information G measure (or the G measure, where G means the generalization of Shannon's information measure) 30 years ago [33], [34], [35], [36]. This measure inherits Carnap and Bar-Hillel's method of using logical probability to measure semantic information, but it is also related to distortion or inaccuracy. Like Shannon's information measure, the G measure also reflects the saved average coding length [35], [36]. In addition, I extended the information rate distortion function $R(D)$ to the information rate fidelity function $R(G)$ [34], [37] ($G$ is the lower limit of average semantic information), which can be used to optimize information efficiency. I use "fidelity" because the amount of semantic information means fidelity (I call $R(G)$ the information rate verisimilitude function in [37], [38] and the rate-of-keeping-precision function in [36], the reasons are similar). In addition, Shannon [1] had used "fidelity" before he discussed distortion. Since semantic information divided by Shannon information represents information efficiency, as well as useful work divided by total work (or energy) means the efficiency of work, $G/R$ means the optimized information efficiency.

In comparison with other semantic or generalized information measures, the G measure is more compatible with Shannon information theory. Moreover, it is consistent with the likelihood method and can be used for machine learning. I have developed some novel algorithms or formulas with the G measure and the $R(G)$ function for maximum mutual information classifications, mixed models [37], Bayesian confirmation [38], [39], and the maximum entropy method [40].

Except for information provided by natural languages, semantic information also includes predictive information and purposive information:

- Predictive information includes information from estimates and observations. This information comes from various objects and is conveyed by instrument readings, economic indicators, senses, perceptions, observations, etc. Observations and descriptions can be regarded as cases of undoubtedly correct predictions. A typical example of predictive information is that provided by a GPS pointer on an electric map. I have discussed predictive information, including sensory information, in my previous articles [34], [36], [37]. However, unlike the predictive information discussed by others [41], [42] what we discuss here is semantic and expressed by truth functions instead of posterior probability distributions.
- Purposive information means purpose-related or goal-related information. This information indicates how well the result (a probability distribution) conforms to the corresponding purpose or goal (a range or fuzzy range). A typical example is the information obtained by the passenger or driver on a car from observing the GPS map displaying the destination and the current position. Another typical example is that the goal is "Grain yields are close to or more than 750 kg/ha (kilogram/hectare)"; we need to evaluate how well the change in the probability distribution of the actual grain yields conforms to this goal. It is the first time I have discussed this information. This semantic information accords with the definition of semantic information in [13]. Different from the constraint function for the constrained optimization in [43], the constraint condition we use is a distribution function (a truth function or likelihood function) instead of some functions' values. The result is also a distribution function (see Section V-B).

This paper will focus on the above two kinds of semantic information. Measuring purposive information requires comparing actual results with expected results. The actual results come from control and observation. Suppose only observation is considered (for example, a passenger observes a GPS map to know how long time it will take to get to the destination). In that case, the system is a communication system. If control is considered simultaneously (for example, the driver can improve the control after observing the GPS map), the system will become a control system. Error or deviation is usually used to express the purposiveness of control, whereas Wiener proposed to use entropy to indicate the uncertainty of control results. We can follow Winner to use Shannon information to indicate the reduced uncertainty by control. However, Shannon entropy or information measure cannot tell how well a control result conforms to a goal (a specified range that may be fuzzy). The semantic information G measure can reflect both error and uncertainty. Measuring purposive information should also be helpful for constrained optimization and reinforcement learning.

The $R(G)$ function reveals an optimal matching relationship between semantic information and Shannon information. I concluded in [37]: we can optimize the semantic channel (or the truth functions) to maximize the amount of semantic information and the information efficiency. This paper will illustrate with an example that we can optimize the Shannon channel to improve the information efficiency (or optimize the control to improve the control efficiency); we can also appropriately sacrifice the efficiency to maximize the purposive information. The information rate fidelity function can help balance the two.

This paper mainly aims at 1) explaining how the amount of semantic information is related to the distortion, freshness, and purposiveness of information; 2) providing a method for maximizing purposive information and information efficiency; 3) hoping that this study will help build a solid bridge between Shannon information theory and the research of semantic communication.

The main contributions of this paper are: 1) enriching semantic information formulas for different scenes; 2) proposing the purpose-related information formulas (for semantic communication and control) and the method of maximizing this information and the information efficiency.

## II. Related Work: Semantic Information G Theory
### A. The P-T Probability Framework

The semantic information G theory is based on the P-T probability framework [37], [38]. This framework includes two types of probabilities: the statistical probability denoted by $P$ and the logical probability by $T$.

*Definition 1:*
- $X$ is a random variable representing an instance, taking a value $x \in U=\{x_1, x_2,...\}$; $Y$ is a random variable denoting a label or hypothesis, taking a value $y \in V=\{y_1, y_2,...\}$.
- The $y_j(x_i)$ is a proposition; $\theta_j$ is a fuzzy subset of $U$, and the elements of $\theta_j$ make $y_j$ true. Then we have $y_j(x) \equiv "x \in \theta_j"$ ("$\equiv$" means they are logically equivalent). The $\theta_j$ may also be a model or a group of model parameters.
- A probability defined with "=", such as $P(y_j) \equiv P(Y=y_j)$, is *a statistical probability*. A probability defined with "$\in$", such as $P(X \in \theta_j)$, is *a logical probability*. To distinguish $P(Y=y_j)$ and $P(X \in \theta_j)$, we define $T(y_j) \equiv T(\theta_j) \equiv P(X \in \theta_j)$ as the logical probability of $y_j$.
- $T(y_j|x) \equiv T(\theta_j|x) \equiv P(X \in \theta_j|X=x)$ is the truth function of $y_j$ and the membership function $m_{\theta_j}(x)$ of $\theta_j$. It changes between 0 and 1.

A semantic channel consists of a group of truth functions: $T(y|x)$: $T(\theta_j|x)$, $j = 1, 2, …, n$. According to the above definition, we have the logical probability:

$$T(y_j) \equiv T(\theta_j) \equiv P(X \in \theta_j) = \sum_i P(x_i)T(\theta_j | x_i). \quad (1)$$

Generally, $T(y_1) + T(y_2) + … + T(y_n) > 1$.

We can put $T(\theta_j|x)$ and $P(x)$ into the Bayes' formula to obtain a likelihood function:

$$P(x|\theta_j) = \frac{T(\theta_j | x)P(x)}{T(\theta_j)}, \quad T(\theta_j) = \sum_i T(\theta_j | x_i)P(x_i). \quad (2)$$

$P(x|\theta_j)$ is called the semantic Bayes' prediction. It is often written as $P(x|y_j, \theta)$ in popular methods. We call the above formula the semantic Bayes' formula. Since the maximum of $T(y|x)$ is 1, from $P(x)$ and $P(x|\theta_j)$, we can obtain:

$$T(\theta_j | x) = \frac{T(\theta_j)P(x|\theta_j)}{P(x)}, \quad T(\theta_j) = 1/\max[P(x|\theta)/P(x)]. \quad (3)$$

Formulas (2) and (3) form the third Bayes' theorem [35] and can be used to convert the likelihood function and the truth function from one to another.

## B. The Semantic Information G Measure

The (amount of) semantic information conveyed by $y_j$ about $x_i$ is defined with log-normalized-likelihood:

$$I(x_i; \theta_j) = \log \frac{P(x_i | \theta_j)}{P(x_i)} = \log \frac{T(\theta_j | x_i)}{T(\theta_j)}. \quad (4)$$

$I(x_i; \theta_j)$, or its average, is the semantic information G measure or the G measure. If $T(\theta_j|x)$ is always 1, the G measure becomes Carnap and Bar-Hillel's semantic information measure [15].

The above formula is illustrated in Figure 1. Figure 1 indicates that the less the logical probability is, the more information there is; the larger the deviation is, the less information there is; a wrong hypothesis conveys negative information. These conclusions conform to Popper's thoughts [44] (p. 294). For this reason, $I(x_i; \theta_j)$ is also explained as verisimilitude [37] between $y_j$ and $x_i$.

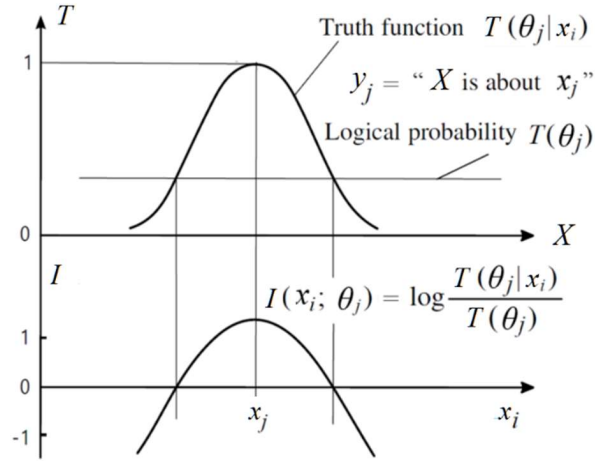

Fig. 1. The semantic information conveyed by $y_j$ about $x_i$ decreases with the deviation or distortion. The larger the deviation is, the less information there is.

We can also use the above formula to measure sensory information, for which $T(\theta_j|x)$ is also understood as the confusion probability or similarity function of $x_j$ with $x$ or the discrimination function of $x_j$ [36].

By averaging $I(x_i; \theta_j)$, we obtain generalized Kullback-Leibler (KL) information or semantic KL information:

$$I(X;\theta_j) = \sum_i P(x_i|y_j)\log\frac{P(x_i|\theta_j)}{P(x_i)} = \sum_i P(x_i|y_j)\log\frac{T(\theta_j|x_i)}{T(\theta_j)}, \quad (5)$$

where $P(x_i|y_j)$, $i = 1, 2, \ldots$, is the sampling distribution, which may be unsmooth or discontinuous. It is easy to prove $I(X; \theta_j) \leq I(X; y_j)$, where $I(X; y_j)$ means the Shannon mutual information $I(X; Y)$ as $Y=y_j$. We can also call it KL information [45]. This form of KL information should belong to Shannon information. Therefore, we also call $I(X; y_j)$ Shannon information hereafter.

When the sample is enormous so that $P(x|y_j)$ is smooth, we may let $P(x|\theta_j) = P(x|y_j)$ or $T(\theta_j|x) \propto P(y_j|x)$ to obtain the optimized truth function [37]:

$$T^*(\theta_j|x) = \frac{P^*(x|\theta_j)}{P(x)} \bigg/ \max\left(\frac{P^*(x|\theta)}{P(x)}\right)$$

$$= \frac{P(x|y_j)}{P(x)} \bigg/ \max\left(\frac{P(x|y)}{P(x)}\right) = \frac{P(x,y_j)}{P(x)P(y_j)} \bigg/ \max\left(\frac{P(x,y)}{P(x)P(y)}\right). \quad (6)$$

According to this equation, $T^*(\theta_j|x_i) = T^*(\theta_{xi}|y_j)$, where $\theta_{xi}$ is a fuzzy subset of $V$. Furthermore, we have

$$T^*(\theta_j|x) \propto P(y_j|x). \quad (7)$$

If $P(x|y_j)$ is unsmooth, we may derive a smooth $T^*(\theta_j|x)$ with parameters by:

$$T^*(\theta_j|x) = \arg\max_{\theta_j} \sum_i P(x_i|y_j)\log\frac{T(\theta_j|x_i)}{T(\theta_j)}. \quad (8)$$

By averaging $I(X; \theta_j)$ for different $y$, we obtain Semantic Mutual Information (SMI):

$$I(X;Y_\theta) = \sum_j P(y_j)\sum_i P(x_i|y_j)\log\frac{P(x_i|\theta_j)}{P(x_i)}$$

$$= \sum_i \sum_j P(x_i)P(y_j|x_i)\log\frac{T(\theta_j|x_i)}{T(\theta_j)} = H(Y_\theta) - H(Y_\theta|X), \quad (9)$$

where

$$H(Y_\theta) = -\sum_j P(y_j)\log T(\theta_j), \quad (10)$$

$$H(Y_\theta|X) = -\sum_j \sum_i P(x_i, y_j)\log T(\theta_j|x_i). \quad (11)$$

$H(Y_\theta)$ can be called the generalized entropy, and $H(Y_\theta|X)$ means the fuzzy entropy [37].

When we fix the Shannon channel $P(y|x)$ and let $P(x|\theta_j) = P(x|y_j)$ or $T(\theta_j|x) \propto P(y_j|x)$ for every $j$, $I(X; Y_\theta)$ reaches its maximum $I(X; Y)$. If we use a group of truth functions or a semantic channel $T(y|x)$ as the constraint function to seek the MMI or maximum information efficiency, we need to let $P(x|y_j) = P(x|\theta_j)$ or $P(y_j|x) \propto T(\theta_j|x)$ as possible for every $j$.

## C. How the Semantic Information G Measure Relates to Distortion

Inspired by the information rate distortion function, I made the following definition in [40].

*Definition 2.* We define *the relationship between the truth function and the distortion function* with:
$$d(x, y) \equiv \log[1/T(y|x)]. \quad (12)$$

Then we have $T(y|x) = \exp[-d(x, y)]$, where we treat log and exp as a pair of function and anti-function (if log means $\log_2$, then $T(y|x) = 2^{-d(x, y)}$). In this way, for any given distortion function $d(x, y_j)$, we can have a semantic information function $I(x; \theta_j)$. For example, letting $d(x, y_j) = (x - x_j)^2 / (2\sigma_j^2)$, we derive:

$$I(x;\theta_j) = \log[1/T(\theta_j)] - (x-x_j)^2/(2\sigma_j^2); \qquad (13)$$

$$I(X;Y_\theta) = -\sum_j P(y_j)\log T(\theta_j) - \sum_j \sum_i P(x_i,y_j)(x_i-x_j)^2/(2\sigma_j^2)$$
$$= I_{max} - \bar{d}, \qquad (14)$$

where $\bar{d}$ is the average distortion, and $I_{max}$ is $H(Y_\theta)$ and the maximum of $I(X; Y_\theta)$ as $\bar{d}$ =0. Because of *Definition 2*, the fuzzy entropy $H(Y_\theta|X)$ becomes the average distortion.

It is easy to find that the semantic information G measure equals Carnap and Bar-Hillel's semantic information measure [15] minus average distortion; $I(X; Y_\theta)$ is like the Regularized Least Square (RLS). Therefore, we can treat the maximum semantic information criterion as a particular RLS criterion. Different from popular RLS criteria, the former only penalizes $\sigma_j$.

### D. Information Rate Fidelity Function R(G)

If we change the distortion limit $\bar{d} \leq D$ for the information rate distortion function into $I(X; Y_\Theta) \geq G$, then the information rate-distortion function becomes the information rate fidelity function [30]. In this case, we replace $d_{ij} = d(x_i, y_j)$ with $I_{ij} = I(x_i; \theta_j)$, like replacing the least square error criterion with the RLS criterion in machine learning. Unlike $d_{ij}$, $I_{ij}$ is related to $P(x)$. The $R(G)$ function pays more attention to $y_j$ with less logical probability than the $R(D)$ function.

Following the derivation of $R(D)$ [3] (p. 31), we obtain the parameterized solution:

$$G(s) = \sum_i \sum_j P(x_i)P(y_j|x_i)I_{ij},$$
$$R(s) = sG(s) - \sum_i P(x_i)\log \lambda_i, \qquad (15)$$

where

$$P(y_j|x_i) = P(y_j)m_{ij}^s/\lambda_i,\ i=1,2,...;\ j=1,2,...$$
$$m_{ij} = T(\theta_j|x_i)/T(\theta_j),\ \lambda_i = \sum_j P(y_j)m_{ij}^s; \qquad (16)$$

We also need the MMI iteration to optimize $P(y)$ and $P(y|x)$ as we do for the information rate distortion function.

The shape of any $R(G)$ function is a bowl-like curve, and its second derivative is greater than 0, as shown in Fig. 2.

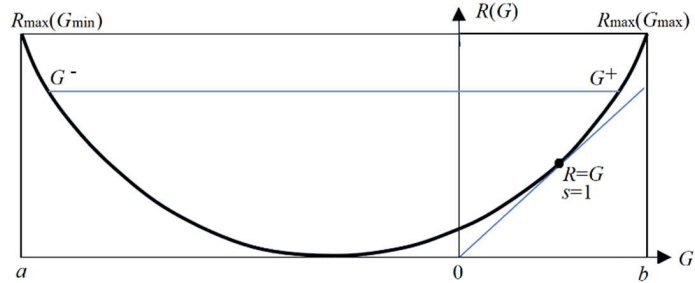

Fig. 2. The information rate fidelity function $R(G)$ for binary communication. For any $R(G)$ function, there is a point where $R(G) = G$. For given $R$, there are two anti-functions $G^-(R)$ and $G^+(R)$.

For the $R(G)$ function, $s=$ d$R$/d$G$. When $s = 1$, $R$ is equal to $G$, which means that the semantic channel matches the Shannon channel. $G/R$ indicates the optimized information efficiency. The $R(G)$ function can be applied to optimize image compression according to visual discrimination [36], semantic compression [40], and the convergence proof of the EM algorithm for mixture models [37], [46].

It is worth noting that for given the semantic channel $T(y|x)$, letting $P(y_j|x) \propto T(y_j|x)$ or $P(x|y_j) = P(x|\theta_j)$ does not maximize SMI but information efficiency. We can maximize SMI by (16) with $s\to\infty$.

We can also replace average distortion $\bar{d}$ with fuzzy entropy $H(Y_\theta|X)$ to obtain the rate-truth function $R(\Theta)$ [40]. $R(G)$ is more suitable than $R(D)$ and $R(\Theta)$ when information is more important than truth. More discussions about the $R(G)$ function's properties can be found in [37], [40].

For given $Y = y_j$ or $P(y_j) = 1$, the $R(G)$ function is also meaningful. In this case, $R$ is the lower limit of $I(X; y_j)$ for given $G = I(X; \theta_j)$; the iteration for $P(y)$ is unnecessary. Section V-B provides such an example.

## III. Predictive Information Related to Inaccuracy and Freshness

### A. Using GPS as an Example to Explain Predictive Information

We consider the information from objects, such as a GPS pointer on a map, a clock, a thermometer, a stock index, a color (or color perception), or a figure formed by many colored dots. For predictive information, $y_j$ becomes the prediction or estimate of $x_j$, i.e., $y_j=\hat{x}_j=$"x is $x_j$".

The truth function $T(y_j|x)$ now is also the confusion probability function or the similarity function of $x_j$ with $x$, which depends on our sensory discrimination or a device's resolution and may be expressed by a negative exponential function. So now the logical probability $T(y_j)$ becomes the confusion probability of $x_j$ with different $x$.

For example, a GPS device's resolution is often represented by Root Mean Square (RMS). Hence, the truth function of $y_j$ or the confusion probability function of $x_j$ is

$$T(y_j | x) = T(\hat{x}_j | x) = \exp[-(x - \hat{x}_j)^2 / (2\sigma^2)]. \quad (17)$$

where $\sigma$ is the RMS. Though $x$ should be two-dimensional, we use one-dimensional $x$ in formulas for convenience. Note that $y_j$ does not determine a probability distribution but a fuzzy range.

*Example 1.* A GPS device is used in a train; hence, $P(x)$ is uniformly distributed on a line (see Fig. 3). The GPS pointer has a deviation. The task is to find the most possible position of the GPS device.

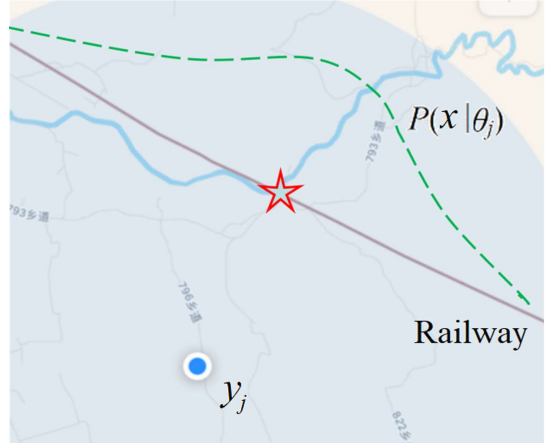

Fig. 3. Illustrating a GPS device's positioning with a deviation. The round point is the pointed position, and the place with the star is the most possible position.

The probability distribution $P(x|\theta_j)$ of $x$ provided by $y_j$ depends on both $P(x)$ and $T(\theta_j|x)$ and can be expressed by the semantic Bayes' formula (2), according to which the star indicates the most possible. If the actual position is $x_i$, the semantic information is $\log[T(\theta_j|x_i)/T(\theta_j)]$. If the actual position is a probability distribution $P(x|y_j) = P(x_i | \hat{x}_j)$, the average semantic information is semantic KL information:

$$I(X;\theta_j) = \sum_i P(x_i | \hat{x}_j) \log \frac{P(x_i | \theta_j)}{P(x_i)} = \sum_i P(x_i | \hat{x}_j) \log \frac{T(\theta_j | x_i)}{T(\theta_j)}. \quad (18)$$

If $P(x_i|\theta_j) = P(x_i | \hat{x}_j)$, then $I(X;\theta_j) = I(X;\hat{x}_j)$, and information efficiency $I(X;\theta_j)/I(X;\hat{x}_j)$ is 1. Otherwise, the information efficiency is less than 1. If the average value of the distribution $P(x|\hat{x}_j)$ is different from $x_j$, or its standard deviation is larger than $\sigma$, which means that the prediction is inaccurate, there will be $I(X;\theta_j) < I(X;\hat{x}_j)$.

Averaging to $I(X; \theta_j)$ for different $y_j$, we can obtain predictive mutual information.

## B. Predictive Information Changes with Time

In (16), we assume there is no time delay between the system's producing $y_j$ and the user's receiving $y_j$. If there is a time delay $\Delta t$, the information will be less. Using Shannon mutual information to evaluate the information freshness has been discussed in [41]. The mutual information is

$$I(X_t; \mathbf{W}_t) = H(X_t) - H(X_t | \mathbf{W}_t), \quad (19)$$

where $\mathbf{W}_t$ denotes the received samples before time $t$. Now we can use the average semantic information to evaluate the information's freshness.

*Example 2.* Consider a GPS pointer's positioning as shown in Fig. 3. Let $t_0=0$, $t_1=\Delta t_0$, and $t = t_1+\Delta t$. For given $y_0 = y(t_0)$ and $y_1 = y(t_1)$, we can make prediction $y(t) = \hat{x}(t)$.

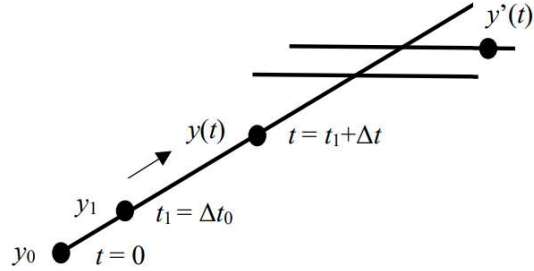

Fig. 3. The GPS pointer's prediction $y(t) = \hat{x}(t)$ changes with $t$. The semantic prediction is made according to $y_0$, $y_1$, and $\Delta t_0$.

Supposing the car's direction and speed are constant, we may predict according to physical knowledge:

$$\hat{x}_1(t) = x_1 + (x_1 - x_0)\Delta t / \Delta t_0, \quad (20)$$

$$T(\hat{x}_1(t) | x) = \exp[-(x - \hat{x}_1(t))^2 / (2\sigma_t^2)], \quad (21)$$

where $\sigma_t$ is greater than $\sigma$, and the longer the $\Delta t$, the larger the $\sigma_t$.

The average semantic information of $\hat{x}_1(t)$ is

$$I(X_t; \theta_1(t)) = \sum_i P(x_i | \hat{x}_1(t)) \log \frac{P(x_i | \theta_1(t))}{P(x_i)} = \sum_i P(x_i | \hat{x}_1(t)) \log \frac{T(\hat{x}_1(t) | x_i)}{T(\hat{x}_1(t))}. \quad (22)$$

$I(X_t;\theta_1(t))$ reaches its upper limit as $P(x|\theta_1(t))=P(x|\hat{x}_1(t))$ for given $P(x|\hat{x}_1(t))$.

Compared with the sampling prediction $P(x|\mathbf{W}_t)$ for $I(X;\mathbf{W}_t)$, the semantic prediction $T(\hat{x}_1(t)|x)$ for $I(X_t;\theta_1(t))$ only makes use of $y_1$, $y_0$, $\Delta t_0$, and $\Delta t$, instead of the samples including many data. The reason is that the GPS device with precision $\sigma$ has been adjusted to fit some samples. Furthermore, the semantic prediction can use the prior knowledge $P(x)$, which is distributed on possible routes.

### C. About Information Freshness

$I(X_t;\theta_1(t))$ will decrease with $\Delta t$ increasing and reflects the information freshness of $y_0$ and $y_1$. The information and freshness will quickly reduce when the car's speed increases and the byroads are closed. The semantic information will become 0 as $\Delta t$ increases to a certain value, denoted by $t_{end}$, which can be treated as the information lifespan of $y_0$ and $y_1$.

If we receive the updated GPS pointer $y=y_j$ at time $t$, then the increment of the semantic information is $I(X;\theta_j)$ - $I(X;\theta_1(t))$. We may decide to update $y$ or not according to the increment. If the car's speed is low and the road has no byroads, the increment should be smaller; hence, we may delay the update. On the other hand, if $y(t)$ becomes $y'(t)$ on a byroad (see Fig. 3), the increment is more prominent, and hence the update in time is necessary.

## IV. PURPOSIVE INFORMATION AND THE INFORMATION EFFICIENCY

### A. Two kinds of Purposive Information

A purpose or goal may be certain or uncertain. In this paper, we consider uncertain purposes. The formulas derived also fit certain purposes. There are two kinds of uncertain purposes[1]: one is expressed by the likelihood function, and another is represented by the truth function, which can be understood as the Distribution Constraint Function (DCF). Since we can set up the relationship between a distortion function and a DCF by letting $d(x,y)=\log[1/T(y|x)]$, the distortion function can be converted into and treated as the DCF.

#### 1) The Generalized KL Information Formula with a Likelihood Function as the Goal

We can use a probability distribution or likelihood function as the goal [37]. For example, a population age distribution, an artillery shell target distribution, or the various chemical elements' proportions in fertilizer can be used as an uncertain goal. The information formula is

$$I(X;c_j/P)=\sum_i P(x_i|\theta_j)\log\frac{P(x_i|c_j)}{P(x_i)}, \qquad (23)$$

where $c_j/P$ means control $c_j$ with $P(x|\theta_j)$ as the goal, and $P(x_i|c_j)$ is the actual distribution. It is easy to prove that $I(X;c_j/P)$ reaches its maximum when we change $P(x|c_j)$ so that $P(x|c_j)=P(x|\theta_j)$.

#### 2) The Semantic KL Information Formula with a Truth Function for Constraint Control

We can also use truth functions as goals or constraints. For example, a truth function indicates

- "The grain yields are close to or more than 750 kg/ha",
- " The wages of workers had better be more than 5000 dollars",
- " The death ages of people had better surpass 70 years old",
- " The cruising distances of the electric vehicles had better reach 500 kilometers", or
- " The error of trains' arrival time should be less than one minute."

---

[1] In my earlier papers (see Sectio 3.3 in [32] and Section 6.4 in [33]), two kinds of goals are mentioned but not well distinguished.

Most of these purposes can be represented by the Logistic function or the negative exponential function. We can express the purposive information by the semantic KL information formula:

$$I(X; c_j / T) = \sum_i P(x_i | c_j) \log \frac{T(\theta_j | x_i)}{T(\theta_j)}, \quad (24)$$

where $c_j/T$ means a control action with a truth function as its purpose. We can obtain the corresponding semantic mutual information by averaging $I(X; c_j/P)$ for different $c_j$.

To ensure that the amount of semantic information is zero when there is no control action so that $P(x|c_j) = P(x)$, we can use the following information increment formula:

$$\Delta I(X; c_j / T) = \sum_i [P(x_i | c_j) - P(x_i)] \log \frac{P(x_i | \theta_j)}{P(x_i)}. \quad (25)$$

## B. Optimizing Purposive Information and Fuzzy Constraint Control

When the actual distribution $P(x|c_j)$ approaches the constrained distribution $P(x|\theta_j)$, the information efficiency, instead of the information, reaches its maximum 1. To further increase $I(X; c_j/T)$, we need to use the SoftMax function, which is used for the $R(G)$ function. From (16), we obtain:

$$P(c_j | x) = P(c_j) m_{ij}^s / \lambda_i, \quad (26)$$

$$P(x_i | c_j) = P(c_j | x_i) P(x_i) / P(c_j) = P(x_i) m_{ij}^s \Big/ \sum_k P(x_k) m_{kj}^s$$
$$= P(x_i)[T(\theta_j | x_i)]^s \Big/ \sum_k P(x_k)[T(\theta_j | x_i)]^s. \quad (27)$$

Since $P(x|c_j)$ above is the function of $\theta_j$ and $s$, we write it as $P(x|\theta_j, s)$ hereafter. It is worth noting that the many distributions of $P(x|c_j)$ meet the constraint and make $I(X; cj/T)$ reach its maximum, but $I(X; c_j)$ is the least as $P(x|c_j) = P(x|\theta_j, s)$.

If $P(x|c_j)$ is obtained from observations, the system is a communication system. On the other hand, if $P(x|c_j)$ is realized by control, the system is a control system. For the latter, the optimization is to improve the control strategy. Therefore, (27) also provides an optimization tool for constraint control. See a computing example in Section V-B.

## V. RESULTS

### A. An Example of Computing the GPS pointer's Predictive Information

I used the following example (illustrated in Fig. 3) to see how the semantic information of a GPS pointer changes with distortion or inaccuracy as time $\Delta t$ increases.

*Example 2 (with details)*. The possible positions of a car form a set $U$, which includes $x_0, x_1, x_2, \ldots, x_{4200}$. $P(x)=1/42000$ is equiprobable on possible roads. The $\sigma$ = RMS is 60. We predict $\hat{x}_1(t)$ according to $x_0 = y_0 = 0$, $t_0 = 0$, $x_1 = y_1 = 100$ (meters), $t_1 = 5$ (seconds), and the assumption that the car's speed and direction are constant. We use $T(\hat{x}_1(t) | x)$ in (19) as the prediction's truth function. We assume that the actual position is also a normal distribution. The task is to answer how semantic information $I(X; \theta_1(t))$ changes with $\Delta t$ or $t = \Delta t + 5$.

Suppose that the actual position's uncertainty and the prediction's inaccuracy increase with $\Delta t$, as shown in TABLE I.

TABLE I. THE PARAMETERS OF THE ACTUAL AND THE PREDICTIVE POSITIONS

|  | Expectation | Standard deviation |
|---|---|---|
| Actual position | $x(t)=100+23\Delta t$ | $\sigma(t)=30+\Delta t$ |
| Inaccurate prediction | $\hat{x}_1(t)=100+20\Delta t$ | $\hat{\sigma}(t)=60+\Delta t$ |
| More accurate prediction | $\hat{x}_1(t)=100+22\Delta t$ | $\hat{\sigma}(t)=60+\Delta t$ |
| Fuzzier prediction | $\hat{x}_1(t)=100+20\Delta t$ | $\hat{\sigma}(t)=60+2\Delta t$ |

I calculated Shannon information $I(X;\hat{x}_1(t))$ for different $\Delta t$ and semantic information $I(X;\theta_1(t))$ for different predictions and $\Delta t$. Fig. 4 shows that Shannon information and semantic information decrease with $\Delta t$ increasing.

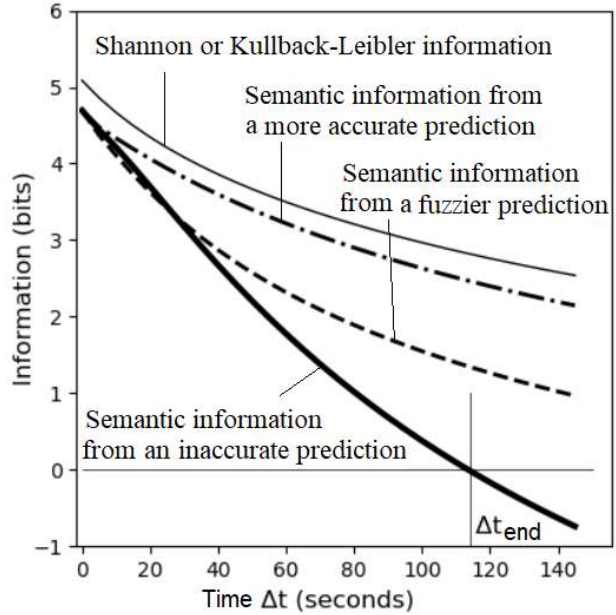

Fig.4. Semantic information and Shannon information decrease with $\Delta t$ increasing. The semantic information $I(X;\theta_1(t))$ for the inaccurate prediction (see the thick solid line) becomes zero as $\Delta t = \Delta t_{end} = 114$ and negative as $\Delta t > 114$. When the prediction becomes more accurate (see the dot-dash line) or fuzzier (see the dashed line), $I(X;\theta_1(t))$ decreases more slowly than before.

Although I could prolong $\Delta t_{end}$ by improving the prediction, the negative semantic information was inevitable with $\Delta t$ increasing.

## C. An Example of Optimizing Purposive Information and the Efficiency

I used the control of adult death ages (by medical conditions) as an example.

*Example 3* (see Fig. 5). Assume that the probability distribution $P(x)$ of current adults' mortality is normal with $\mu = 50$ and $\sigma = 10$. A constraint function represents the goal:

$$T(\theta_j | x) = 1/[1+\exp(-0.8(x-60))], \tag{28}$$

The task is to find the control result' distribution $P(x|c_j)$ for the highest information efficiency and the maximum semantic information $I(X;\theta_j/T)$ with as less Shannon information $I(X;c_j)$ as possible.

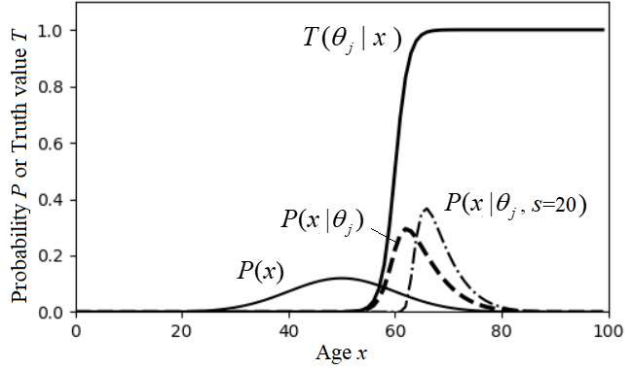

Fig. 5. The constraint function $T(\theta_j|x)$ represents the goal, and the likelihood functions denote the results of control. $P(x|\theta_j) = P(x|\theta_j, s=1)$ (see the dashed line) is for $G = R$, and $P(x|\theta_j, s = 20)$ (see the dot-dash line) makes $G$ approach its maximum.

For this example, $R$ denotes the lower limit of $I(X; c_j)$ for given $G = I(X; \theta_j/T)$, and $G$ represents the upper limit of $I(X; \theta_j/T)$ for given $R = I(X; c_j)$.

I obtained $R(G)$ by using (27) and different $s$, as shown in Fig. 6. The average distortion $\bar{d}$ is also the function of $s$, i.e.,

$$\bar{d}(s) = -\sum_i P(x_i|c_j,s)\log T(\theta_j|x_i). \tag{29}$$

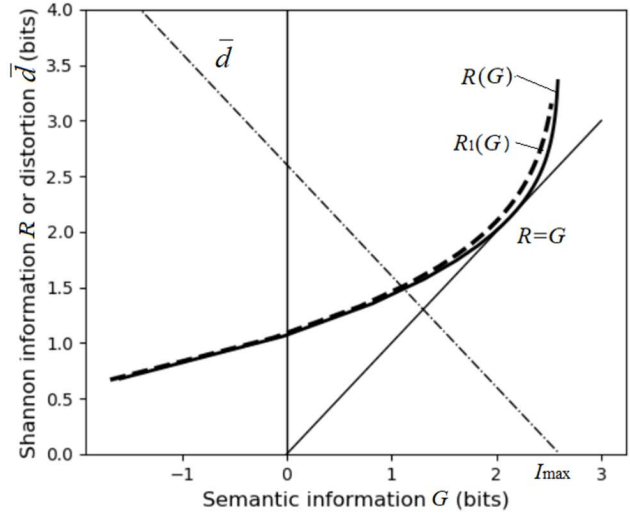

Fig. 6. Two functions: $R(G)$ (solid line) was produced by using $P(x|c_j)=P(x|\theta_j, s)$; $R_1(G)$ (dashed line) was made by using $P(x|c_j) = P(x|\beta_j, s)$, which is a normal distribution; $\bar{d}$ (dot-dash line) is the average distortion; $I_{max} = \log[1/T(\theta_j)] = G + \bar{d}$.

In this case, $I_{max} = \bar{d} + G$ is constant. If we calculate the average semantic information of $c_1, c_2, \ldots, I_{max}$ will change with probability distribution $P(c)$.

The results (in Fig. 6 and TABLE II) illustrate that information efficiency $R/G$ reaches its maximum, 1, as $P(x|c_j) = P(x|\theta_j, s = 1) = P(x|\theta_j)$. Shannon information $R$ increases faster than semantic information $G(R)$ as $s > 1$. $G(R)$ almost stops at 2.58 bits as $s > 20$. The $R(G)$ function reminds that we need a trade-off between maximizing semantic information and maximizing information efficiency. In this case, $s \approx 5$ should be a good choice.

If $P(x|c_j)$ was produced by some parameters, $R(G)$ became $R_1(G)$. To see the difference, I obtained $R_1(G)$ by using $P(x|c_j) = P(x|\beta_j, s)$, a normal distribution. I used the expectation and square error of $P(x|\theta_j, s)$ for $P(x|\beta_j, s)$. TABLE II shows $R(G)$ and $R_1(G)$ and the efficiencies changing with $s$.

TABLE II. $R(G)$, $R_1(G)$, AND THE EFFICIENCIES CHANGE WITH $S$

| Function | Infor. (bits) | $s = 1$ | $s = 20$ | $s = 40$ |
|---|---|---|---|---|
| $R(G)$ | $R$ | 2.19 | 3.36 | 3.58 |
|  | $G$ | 2.19 | 2.58 | 2.59 |
|  | $G/R$ | 1 | 0.77 | 0.72 |
| $R_1(G)$ | $R_1$ | 2.08 | 3.13 | 3.38 |
|  | $G$ | 1.99 | 2.52 | 2.55 |
|  | $G/R_1$ | 0.95 | 0.80 | 0.76 |

There is always $R_1 < R$ for any given $G$, which means $R$ has a higher efficiency than $R_1$. This is because $R_1(G)$ cannot pass through the point at which $R = G$ or $G/R = 1$. After all, $P(x|\beta_j)$ is a symmetrical normal distribution, whereas $P(x|\theta_j)$ is asymmetrical (see Fig. 5). However, $G_1/R_1 = 76$ for $s = 40$ is similar to $G/R=0.77$ for $s = 20$. The similar values mean that the normal function is good enough for this example if we select a suitable $s$.

One may think that we can select $x$, such as $x = 80$, that maximizes $\log[T(\theta_j|x)/T(\theta_j)]$; $P(x=80|c_j) = 1$ is what we seek. However, this selection cannot guarantee the information efficiency is high enough. To test this conclusion, I replaced $P(x|c_j)$ with $P(x = 80|c_j)=1$ to obtain the following results. The semantic information is $\log[1/T(\theta_j)] = 2.60$ bits; the Shannon information is $\log[1/P(x=80)] = 11.14$ bits; the information efficiency is 0.23, which is much lower than $G/R$ and $G/R_1$ in TABLE II.

## VI. Discussion
### A. About the Distortion and Freshness of Information

In Shannon information theory, more mutual information does not guarantee less distortion or higher accuracy. When estimates and predictions are inaccurate, mutual information measure is not suitable for measuring information received and understood by people. Therefore, we need a semantic information measure that reflects both inaccuracy and reduced coding length. Figs. 1, 3, and 4 support the conclusion that the semantic information G measure is such a measure.

Fig. 4 shows that semantic information decreases with accuracy decreasing because of time and the parameters. Unlike Shannon information, semantic information may easily become 0 (at $\Delta t_{end}$) and negative (as $\Delta t > \Delta t_{end}$). Therefore, we may consider using $t_{end}$ as the message's information lifespan and defining

$$r_t = [1-I(X; \theta_1(t)) / I(X; \theta_1(t_1))]*100\% \tag{30}$$

as the relative age of information for the message's update.

However, this paper discusses information freshness only with a simple example related to GPS. Generally, semantic predictions are complicated; the semantic Bayes' formula is helpful but not enough. Therefore, it is necessary to combine the semantic information methods in this paper with existing methods.

## B. About Purposive Information for Constraint Control

The mean square error is often used to evaluate whether control results meet purposes. Weiner proposes to use entropy to indicate the uncertainty of control results. According to Weiner's idea, Shannon information means control complexity. However, using entropy or Shannon information is not convenient for expressing the goal of control and how the actual results conform to the goal. Section IV-A shows that we can use the goal-related information formula to describe the purposiveness.

Furthermore, Section IV-B and Section V-B demonstrate that we can optimize the constraint control with the help of the $R(G)$ function. Section V-B provides an example for maximizing the purposive information and information efficiency $G/R$. This example can help us balance the amount of semantic information (or control) and the efficiency of information (or control).

However, the application of the $R(G)$ function to source coding still needs further studies.

## C. The Various Semantic Information Formulas for the G Measure

TABLE II, as a summary, lists various semantic information formulas that express G measure for different scenes.

TABLE II. THE SIMILARITIES AND DIFFERENCES BETWEEN VARIOUS SEMANTIC INFORMATION FORMULAS

| Information type | Example of conveying information | Learning/ constraint function | Maximizing information by | Distortion (function) may be | Average semantic information formula |
|---|---|---|---|---|---|
| Natural languages' semantic information | "It will be heavily rainy tomorrow." | Truth or membership function | Change $\theta_j$ for $P(x\|\theta_j) = P(x\|y_j)$ or $T(\theta_j\|x) \propto P(y_j\|x)$ | $d(y, x) = \log[1/T(y\|x)]$ | $I(X;\theta_j) = \sum_i P(x_i\|y_j)\log\dfrac{T(\theta_j\|x_i)}{T(\theta_j)}$ |
| Semantic predictive information | A GPS pointer on an electronic map | Truth or confusion prob. function | As above | $d(x, y) = (x - y)^2/(2\sigma^2)$ | $I(X;\theta_j) = \sum_i P(x_i\|\hat{x}_j)\log\dfrac{T(\theta_j\|x_i)}{T(\theta_j)}$ |
| Purposive information $I(X; c_j/T)$ | The goal is to make grain yields close to or more than 7000 kg/ha | Truth or membership function | Change $c_j$ for $P(x\|c_j)=P(x\|\theta_j)$, then use SoftMax func. | $d(y, x) = \log[1/T(y\|x)]$ | $I(X;c_j/T) = \sum_i P(x_i\|c_j)\log\dfrac{T(\theta_j\|x_i)}{T(\theta_j)}$ |
| Probability prediction information | "The probability of rain in the afternoon is 0.3." | Likelihood function | Change $\theta_j$ for $P(x\|\theta_j) =P(x\|y_j)$ | KL divergence | $I(X;\theta_j) = \sum_i P(x_i\|y_j)\log\dfrac{P(x_i\|\theta_j)}{P(x_i)}$ |
| Purposive information $I(X; c_j/P)$ | The artillery's target is a normal distribution. | Likelihood function | Change $c_j$ for $P(x\|c_j) = P(x\|\theta_j)$ | KL divergence | $I(X;c_j/P) = \sum_i P(x_i\|\theta_j)\log\dfrac{P(x_i\|c_j)}{P(x_i)}$ |

Note that optimizing methods may be different for different scenes or formulas.

## D. Similar Information Measures in Deep Learning

The InforNCE loss function [48], [49] is very successful for deep learning. It is written as [49]:

$$L_q = -\log\frac{\exp(q.k_+/\tau)}{\sum_{i=0}^{K}\exp(q.k_i/\tau)}, \qquad (31)$$

where $L_q$ means the information loss of treating vector $q$ as vector $k_+$ ($k_+$ is one of $k_0, k_1, \ldots$), the dot product $q.k_+$ means the similarity between $q$ and $k_+$, and $\tau$ means "temperature". Now we define the distortion function between $q$ and $k_i$ by $d(q, k_i)=(m - q.k_i)/\tau$, where $m$ is the maximum of $q.k_i$, $i=1,2,\ldots$ . Then we have

$$L_q = -\log \frac{\exp(q.k_+/\tau)/\exp(m/\tau)}{\sum_{i=0}^{K}\exp(q.k_i/\tau)/\exp(m/\tau)} = -\log \frac{\exp[-d(q,k_+)]}{\sum_{i}^{K}\exp[-d(q,k_i)]} = -I(q;k_+), \tag{32}$$

where $I(q; k_+)$ can be regarded as the semantic information conveyed by $k_+$ about $q$.

The neural information measure $I_\Theta(X; Z)$ in [50] is also similar to the G measure. Furthermore, the $R(G)$ function is related to the Expectation-Maximization algorithm [46], the Restricted Boltzmann Machine [40], some Minimization-Maximization algorithms, and the SoftMax function [40].

The above facts indicate that the semantic information G measure is compatible with some advanced deep learning methods. However, we need further studies to apply this theory, especially the method in which the semantic channel and Shannon's channel mutually match for maximum semantic information or maximum information efficiency, to deep learning with neural networks.

### E. The Optimization of Semantic Communication for Information Values

Communication efficiency includes information efficiency but may also involve utility or information value. Although "Information Value" has some different meanings, such as in [10] (Section 6.2), [11], [50], its primary meaning should be the increment of utility or wealth because of information.

Cover and Thomas in [10] discussed the information value by combining Shannon information theory and Portfolio Compound Interest Theory (PCIT). I further discussed the information value by combining the semantic information G measure and the PCIT, where I provided some useful formulas, including several generalized Kelly formulas [50]. We can also extend the $R(G)$ function to the $R(V)$ function ($V$ is the lower limit of the average value of semantic information; $I_{ij}$ becomes $v_{ij}$ [34]) for optimizing Shannon channel $P(y|x)$.

However, we still need further studies to optimize semantic communication for general information values, such as those related to weather forecasts, disease control, and wars.

## REFERENCES

[1] C. E. Shannon, "A Mathematical Theory of Communication," *Bell System Technical Journal*, vol. 27, no. 3, pp. 379–423, 1948, doi: 10.1002/J.1538-7305.1948.TB01338.X.

[2] C.E. Shannon, "Coding theorems for a discrete source with a fidelity criterion." *IRE Nat. Conv. Rec.*, Part 4, 142–163, 1959.

[3] T. Berger, "Rate Distortion Theory." Prentice-Hall, Englewood Cliffs, New Jersey, USA, 1971, pp. 30-32.

[4] J. Gibson, "Special issue on rate distortion theory and information theory," *Entropy*, vol. 20, no. 11, Nov. 2018, doi: 10.3390/E20110825.

[5] S. Kaul, R. Yates, and M. Gruteser, "Real-time status: How often should one update?" *Proceedings - IEEE INFOCOM*, pp. 2731–2735, 2012, doi: 10.1109/INFCOM.2012.6195689.

[6] Y. Sun, E. Uysal-Biyikoglu, R. Yates, C. E. Koksal, and N. B. Shroff, "Update or wait: How to keep your data fresh," *Proceedings - IEEE INFOCOM*, vol. 2016-July, Jul. 2016, doi: 10.1109/INFOCOM.2016.7524524.

[7] A. Kosta, N. Pappas, and V. Angelakis, "Age of Information: A New Concept, Metric, and Tool," *Foundations and Trends in Networking*, vol. 12, no. 3, pp. 162–259, 2017, doi: 10.1561/1300000060.

[8] M. Bastopcu and S. Ulukus, "Age of Information for Updates with Distortion: Constant and Age-Dependent Distortion Constraints," *IEEE/ACM Transactions on Networking*, vol. 29, no. 6, pp. 2425–2438, Dec. 2021, doi: 10.1109/TNET.2021.3091493.

[9] W. Weaver and C. Shannon, "Recent Contributions to The Mathematical Theory of Communication," *Zach. f. Phys*, vol. 53, 1925.

[10] Cover, T. M.; Thomas, J. A. Elements of Information Theory, John Wiley & Sons: New York, USA, 2006.

[11] I. S. Moskowitz, S. Russell, and W. F. Lawless, "An information geometric look at the valuing of information," *Human-Machine*

[12] E. Calvanese Strinati and S. Barbarossa, "6G networks: Beyond Shannon towards semantic and goal-oriented communications," *Computer Networks*, vol. 190, May 2021, doi: 10.1016/J.COMNET.2021.107930.

[13] E. Uysal *et al.*, "Semantic Communications in Networked Systems: A Data Significance Perspective," Mar. 2021, Accessed: Dec. 14, 2022. [Online]. Available: http://arxiv.org/abs/2103.05391

[14] J. Bao *et al.*, "Towards a theory of semantic communication," *Proceedings of the 2011 IEEE 1st International Network Science Workshop, NSW 2011*, pp. 110–117, 2011, doi: 10.1109/NSW.2011.6004632.

[15] R. Carnap and Y. Bar-Hillel, *An outline of a theory of semantic information*, vol. 19, no. 3. Research laboratory of electronics, Massachusetts institute of technology, 1954. doi: 10.2307/2268645.

[16] L. Floridi, "Outline of a Theory of Truth as Correctness for Semantic Information," *tripleC: Communication, Capitalism & Critique. Open Access Journal for a Global Sustainable Information Society*, vol. 7, no. 2, pp. 142–157, Nov. 2009, doi: 10.31269/TRIPLEC.V7I2.131.

[17] T. Hosseini and M. Jabbari Nooghabi, "Discussion about inaccuracy measure in information theory using co-copula and copula dual functions," *J Multivar Anal*, vol. 183, May 2021, doi: 10.1016/J.JMVA.2021.104725.

[18] P. Afshar, A. Nobakhti, H. Wang, and T. Chai, "Multi-objective minimum entropy controller design for stochastic processes," *Proceedings of the 2010 American Control Conference, ACC 2010*, pp. 355–360, 2010, doi: 10.1109/ACC.2010.5530823.

[19] M. Ren, Q. Zhang, and J. Zhang, "An introductory survey of probability density function control," *http://mc.manuscriptcentral.com/tssc*, vol. 7, no. 1, pp. 158–170, Jan. 2019, doi: 10.1080/21642583.2019.1588804.

[20] X. Xu and S.-L. Huang, "On Distributed Learning With Constant Communication Bits," *IEEE Journal on Selected Areas in Information Theory*, vol. 3, no. 1, pp. 125–134, Mar. 2022, doi: 10.1109/JSAIT.2022.3157797.

[21] R. Kamimura, "Cost-conscious mutual information maximization for improving collective interpretation of multi-layered neural networks," *Neurocomputing*, vol. 409, pp. 259–274, Oct. 2020, doi: 10.1016/J.NEUCOM.2020.04.127.

[22] S. Jain, M. Cardone, and S. Mohajer, "Optimality of energy efficient scheduling and relaying for half-duplex relay networks," IEEE J. Sel. Areas Inf. Theory, vol. 3, no. 1, Mar. 2022, doi: 10.1109/JSAIT.2022.3157829

[23] E. Bonnevie, "Dretske's semantic information theory and meta-theories in library and information science," *Journal of Documentation*, vol. 57, no. 4, pp. 519–534, 2001, doi: 10.1108/EUM0000000007093.

[24] L. Floridi, "Outline of a Theory of Strongly Semantic Information," *Minds Mach (Dordr)*, vol. 14, no. 2, pp. 197–221, May 2004, doi: 10.1023/B: MIND.0000021684.50925.C9.

[25] S. D'alfonso, "On quantifying semantic information," *Information (Switzerland)*, vol. 2, no. 1, pp. 61–101, Mar. 2011, doi: 10.3390/INFO2010061.

[26] Y. Zhong, "A theory of semantic information," *China Communications*, vol. 14, no. 1, pp. 1–17, Jan. 2017, doi: 10.1109/CC.2017.7839754.

[27] J. Yang, Y. Li, C. Gao, and Y. Zhang, "Measuring the short text similarity based on semantic and syntactic information," *Future Generation Computer Systems*, vol. 114, pp. 169–180, Jan. 2021, doi: 10.1016/J.FUTURE.2020.07.043.

[28] G. J. Klir, "Generalized information theory," *Fuzzy Sets Syst*, vol. 40, no. 1, pp. 127–142, Mar. 1991, doi: 10.1016/0165-0114(91)90049-V.

[29] I. S. Moskowitz, S. Russell, and W. F. Lawless, "An information geometric look at the valuing of information," *Human-Machine Shared Contexts*, pp. 177–204, Jan. 2020, doi: 10.1016/B978-0-12-820543-3.00009-2.

[30] A. de Luca and S. Termini, "A definition of a nonprobabilistic entropy in the setting of fuzzy sets theory," *Information and Control*, vol. 20, no. 4, pp. 301–312, 1972, doi: 10.1016/S0019-9958(72)90199-4.

[31] D. Bhandari and N. R. Pal, "Some new information measures for fuzzy sets," *Inf Sci (N Y)*, vol. 67, no. 3, pp. 209–228, Jan. 1993,

doi: 10.1016/0020-0255(93)90073-U.

[32] K. Tanuj, K. B. Rakesh, and G. Nitin, "On Some Parametric Generalized Measures of Fuzzy Information, Directed Divergence and Information Improvement," *Int J Comput Appl*, vol. 30, no. 9, pp. 5–10, 2011, doi: 10.5120/3666-5185.

[33] C. Lu, "Shannon equations reform and applications," *BUSEFAL*, vol. 44, no. 4, 45-52, 1990. https://www.listic.univ-smb.fr/production-scientifique/revue-busefal/version-electronique/ebusefal-44/

[34] C. Lu, A Generalized Information Theory (广义信息论), China Science and Technology University Press (中国科学技术大学出版社): Hefei, China, 1993. ISBN 7-312-00501-2.

[35] C. Lu, "Meanings of generalized entropy and generalized mutual information for coding (广义熵和广义互信息的编码意义)," *J. of China Institute of Communication* (《通信学报》), vol. 5, no. 6, 37-44, June 1994.

[36] C. G. Lu, "A generalization of Shannon's information theory," *International Journal Of General System*, vol. 28, no. 6, pp. 453–490, 1999, [Online]. Available: http://www.tandfonline.com/doi/abs/10.1080/03081079908935247

[37] C. Lu, "Semantic information G theory and logical Bayesian inference for machine learning," *Information (Switzerland)*, vol. 10, no. 8, 2019, doi: 10.3390/info10080261.

[38] C. Lu, "The P–T Probability Framework for Semantic Communication, Falsification, Confirmation, and Bayesian Reasoning," *Philosophies*, vol. 5, no. 4, p. 25, Oct. 2020, doi: 10.3390/philosophies5040025.

[39] C. Lu, "Channels' confirmation and predictions' confirmation: From the medical test to the Raven Paradox," *Entropy*, vol. 22, no. 4, Apr. 2020, doi: 10.3390/E22040384.

[40] C. Lu, "Using the Semantic Information G Measure to Explain and Extend Rate-Distortion Functions and Maximum Entropy Distributions," *Entropy*, vol. 23, no. 8, Aug. 2021, doi: 10.3390/E23081050.

[41] Y. Sun and B. Cyr, "Information Aging Through Queues: A Mutual Information Perspective," *IEEE Workshop on Signal Processing Advances in Wireless Communications, SPAWC*, vol. 2018-June, Aug. 2018, doi: 10.1109/SPAWC.2018.8445873.

[42] S. E. Palmer, O. Marre, M. J. Berry, and W. Bialek, "Predictive information in a sensory population," *Proc Natl Acad Sci U S A*, vol. 112, no. 22, pp. 6908–6913, Jun. 2015, doi: 10.1073/PNAS.1506855112/SUPPL_FILE/PNAS.201506855SI.PDF.

[43] J. Gregory and C. Lin, "Constrained optimization in the calculus of variations and optimal control theory," *Constrained Optimization in the Calculus of Variations and Optimal Control Theory*, pp. 1–217, Jan. 2018, doi: 10.1201/9781351070867/CONSTRAINED-OPTIMIZATION-CALCULUS-VARIATIONS-OPTIMAL-CONTROL-THEORY-GREGORY.

[44] K. R. Popper, *Conjectures and refutations : the growth of scientific knowledge*. Routledge, 2002. Accessed: Dec. 14, 2022. [Online]. Available: https://www.routledge.com/Conjectures-and-Refutations-The-Growth-of-Scientific-Knowledge/Popper/p/book/9780415285940

[45] S. Kullback and R. A. Leibler, "On Information and Sufficiency," vol. 22, no. 1, pp. 79–86, Mar. 1951, doi: 10.1214/AOMS/1177729694.

[46] C. Lu, "Understanding and Accelerating EM Algorithm's Convergence by Fair Competition Principle and Rate-Verisimilitude Function," Apr. 2021, doi: 10.48550/arxiv.2104.12592.

[47] A. van den Oord DeepMind, Y. Li DeepMind, and O. Vinyals DeepMind, "Representation Learning with Contrastive Predictive Coding," Jul. 2018, doi: 10.48550/arxiv.1807.03748.

[48] K. He, H. Fan, Y. Wu, S. Xie, and R. Girshick, "Momentum Contrast for Unsupervised Visual Representation Learning," Proceedings of the IEEE Computer Society Conference on Computer Vision and Pattern Recognition, pp. 9726–9735, Nov. 2019, doi: 10.48550/arxiv.1911.05722.

[49] M. I. Belghazi et al., "MINE: Mutual Information Neural Estimation," pp. 1–44, Jan. 2018, doi: 10.48550/arxiv.1801.04062.

[50] C. Lu, C. *The Entropy theory of portfolio and information value: on the risk control of stocks and futures* (投资组合的熵理论和

*信息价值——兼谈股票期货风险控制*). Science and Technology University Press (中国科学技术大学出版社): Hefei, China, 1997， ISBN7-312-00952-2/F.36. http://survivor99.com/lcg/books/portfolio/index.htm

The author email： survival99@gmail.com
Homepage: http:// Chenguang Lu first page (survivor99.com)